\documentclass[aip,jcp,twocolumn,reprint,amsfonts,amssymb,graphicx,epsfig]{revtex4-1}
\usepackage{amssymb}
\usepackage[dvips]{graphicx}
\usepackage{epsfig}
\usepackage{graphicx}

\begin{document}

\title{Melting slope of MgO from molecular dynamics and density functional theory\\}

\author{Paul Tangney$^{1}$ and Sandro Scandolo$^{2}$
} 

\affiliation{
$^1$ Department of Physics and Department of Materials, Imperial College London, London SW7 2AZ, United Kingdom.
\\
$^2$ The Abdus Salam International Centre for Theoretical Physics (ICTP)
     and INFM/Democritos National Simulation Center,
     Strada Costiera 11, I-34014 Trieste, Italy.
\\
}

\date{\today}

\begin{abstract}
We combine density functional theory (DFT) with molecular dynamics simulations based on an 
accurate atomistic force-field to calculate the pressure derivative of the melting 
temperature of magnesium oxide
at ambient pressure - a quantity for which a serious disagreement between theory 
and experiment has existed for almost 15 years.
We find reasonable agreement with previous DFT results and with a very recent experimental 
determination of the slope.
We pay particular attention to areas of possible weakness in theoretical calculations and conclude
that the longstanding discrepancy with experiment could only be explained by a dramatic failure of existing
density functionals or by flaws in the original experiment.
\end{abstract}
\vspace{0.5cm}

\maketitle
\section{Introduction}
MgO is a major component of the Earth's mantle and so its thermodynamic properties
at high pressures (P) and temperatures (T) are crucial to our understanding of its composition and 
evolution.
It is arguably the simplest oxide\cite{cohen}, being
stable in the NaCl cubic structure at pressures up to at least 227 GPa at ambient temperature\cite{duffy_mgo_eos},
and its simplicity and abundance
make it a natural starting point for attempts to understand and model oxides of geophysical
relevance~\cite{cohen_and_weitz}.

Computer simulations based either on quantum mechanics or on atomistic force-fields are playing 
an increasingly important role in geophysical research because the
experimental difficulties at the extreme temperatures and pressures relevant 
to the Earth's mantle are considerable.
However, for simulations to be of use, it is important to be able to rely on the accuracy of theoretical 
descriptions of interactions between atoms and to understand their limitations.
When bonding can be described accurately and efficiently, simulations
allow many physical properties of materials to be calculated at arbitrary temperatures and
pressures. 
On the other hand, an inability to accurately calculate the properties
of an oxide as simple as MgO would cast serious doubt on the suitability of computer simulations for quantitative
studies of more complicated oxides such as (Mg,Fe)SiO$_3$-perovskite and (Mg,Fe)O magnesiow\"{u}stite, which together
make up about 90\% of the lower mantle.

Both MgO and MgSiO$_3$ 
are known to melt at temperatures substantially 
above the geothermal profile, however, a quantitative determination of 
their melting temperatures ($T_m$) at high pressures is a crucial parameter in
rheological, geodynamical, and chemical differentiation models of the lower 
mantle~\cite{yamazaki,boehler_review}. 
Viscosity models, for example, scale with the ``homologous'' temperature 
($T/T_m$), $T$ being the actual temperature along the geotherm~\cite{weertman}.
Chemical differentiation in the early, partially molten state 
of the mantle must have occurred at temperatures above the MgSiO$_3$/MgO 
solidus, which is in turn determined by $T_m$ of the end-members.

Until recently, only one experimental measurement of the melting temperature of
MgO at high pressure existed\cite{zerr_and_boehler} and this extrapolated
to a rather low value of 5000 K for the melting temperature of MgO
at core-mantle boundary pressures (130 GPa). If correct, this would imply that 
viscosity in the lower mantle is dominated by atomic diffusion in MgO
~\cite{yamazaki} and suggest that partial melting may be the cause of the 
seismic anomalies at the bottom of the mantle~\cite{boehler_review}.
However, atomistic modeling has consistently yielded a much steeper increase ($dT_m/dP$)
of the melting temperature of MgO with pressure
\cite{cohen_and_gong,belonoshko,vocadlo_and_price,cohen_and_weitz,strachan,alfe_mgo,madden}
The theoretical estimates of $dT_m/dP$ range from 
88 K/GPa \cite{strachan} to 270 K/GPa \cite{cohen_and_gong}, while Zerr and Boehler found a value
of 36 K/GPa~\cite{zerr_and_boehler}.

Because the melting slope is related through the Clapeyron relation 
($dT_m/dP = T_m \Delta V / \Delta E$) to fundamental physical properties of the material such as the change in 
molar volume upon melting ($\Delta V$) and the latent heat $\Delta E$, if the results of Zerr and Boehler were
correct, it would point to a dramatic failure of atomistic models. However, even at low pressures, 
there are considerable difficulties associated with experimental measurements of the MgO melting point\cite{ronchi}. 
Ronchi and Sheindlin
have reported a zero pressure melting point of $3250\pm 20$ K\cite{ronchi} which differs significantly from the value of 
$3040\pm 100$ measured by Zerr and Boehler. Further doubt has been cast on Zerr and Boehler's measurements by
a very recent experiment in which 
Zhang and Fei have extrapolated a value of $dT_m/dP = 221$ K/GPa from 
measurements of melting of (Mg,Fe)O solid solutions at high pressure\cite{zhang_and_fei}.

Here we combine molecular dynamics simulations with density functional theory (DFT)
to determine the melting slope of MgO. There have been two previous calculations of the 
melting slope that relied heavily on DFT - one by Alf\'e\cite{alfe_mgo}
and one by Aguado and Madden\cite{madden}. Our calculations are intended to complement these studies and to 
demonstrate that, relative to the large discrepancy between the calculated melting slope and 
that measured by Zerr and Boehler, there is agreement between values of the melting slope 
calculated by different groups using DFT. 
The calculations are also in better agreement with the work of Zhang and Fei.

We pay particular attention
to analysing possible sources of error in our calculations 
and in previous calculations. 
Our calculations indicate 
that the rate of increase of the melting 
temperature with pressure is between three and 
five times steeper than reported experimentally by
Zerr and Boehler. Furthermore, 
this discrepancy does not appear to be explainable by statistical uncertainties
in our calculations or by differences in the description of interatomic forces between the
model potentials that we use for efficient statistical sampling and DFT.
We are forced to conclude that either the local and generalised-gradient approximations to DFT 
fail spectacularly for solid and/or liquid MgO, or
there are problems with the experimental results of Zerr and Boehler.
Such a dramatic failure of density functionals for a material as simple as MgO would be very surprising and, to our knowledge, unprecedented.
Therefore, given that the experiment of Zhang and Fei casts doubt on the results of Zerr and Boehler, it seems likely that 
current models of viscosity for the Earth's mantle which rely on these results need to be revised.

\section{Previous calculations}
Many of the early calculations of the melting slope were performed using
empirical atomistic models. These are energy functions of the atomic coordinates
that were parametrized using low-temperature experimental or DFT data for crystalline MgO. 
There are several potential problems with these calculations. One potential problem is that many of
these atomistic models do not adequately describe electronic effects, such as ionic
polarization, that may have a significant impact on thermodynamics. This is of particular
concern for pairwise-additive force-fields which do not contain any phenomenological representation
of the response of electrons on an ion to changes in the ion's environment.
A second potential problem is that the data to which these atomistic models 
were fit does not relate directly to the relevant thermodynamic $(P,T)$ conditions and it doesn't relate directly
to the liquid state. This means that one can be less confident of the models' applicability under these 
conditions where disorder and changes in volume may alter the electronic structure.
Finally, if the quantity of data to which a model is fit is small it is relatively easy to achieve
a good fit. However, one can never be sure that this fit results in a good underlying description
of the forces on the ions.
For these reasons, it is difficult to assess
the reliability of calculations of the melting slope that are based on these purely-empirical 
atomistic models.

Parameter-free (or first-principles) approaches based on density functional 
theory and on the full description of the quantum electronic ground state
have proven to be much more accurate and reliable than conventional force-fields for 
the calculation of the static and vibrational properties of crystalline MgO 
at low temperature\cite{karki}. A serious drawback of first-principles approaches, however, 
is their computational expense, which limits simulations to short time and length scales.
Therefore, statistical sampling is usually poor 
and the precision with which thermodynamic properties can be calculated is low.
Nevertheless, recent methodological advances and increasing computational 
resources have allowed the study of high-T thermodynamic properties of minerals 
in a few cases,
including melting~\cite{alfe_nature} and thermoelasticity~\cite{oganov_nature}.
To find the reason for the discrepancy between theory and experiment on 
the melting slope of MgO, we will attempt to rule out as many of the possible reasons
for this discrepancy as we can.
Because simulations that rely solely on empirical or semi-empirical atomistic force-fields yield 
calculated melting slopes that differ by 
up to a factor of three and because their accuracy is very difficult to assess,
we must assume, for the sake of the present argument, that they are untrustworthy.
Therefore, we consider only the more recent calculations of $dT_m/dP$ that have
been performed with substantial help from first-principles calculations.

Alf\'{e} has calculated the melting slope of MgO without any reliance
on atomistic force-fields by performing first principles molecular dynamics. 
He found a value for the melting slope of $102 \pm 5$ K/GPa\cite{alfe_mgo}.
However, we cannot rule out the possibility that his results are affected by 
uncertainties arising from short equilibration times and production runs.
Aguado and Madden, on the other hand, have substantially reduced the probability
of poor equilibration and substantially increased the precision with which thermodynamic
properties are calculated by using a highly-accurate atomistic potential that
has been parametrized using DFT \cite{madden}. Equilibration and statistical sampling
have been performed with this relatively-efficient force-field and they have used
DFT to check the accuracy of the calculated energy differences between solid 
and liquid. They find a melting slope of $125$ K/GPa.
Although the force-field that they use is very good, the configurations that they generate
to calculate energy and volume differences
can not be trusted as much as those that would be calculated if dynamics had been performed on 
the DFT potential energy surface.
For example, they have parametrized their potential by fitting to DFT calculations 
of configurations from a number of different solid phases. 
There is no guarantee that their force-field would be transferable to the liquid if the liquid structure 
differed strongly from these crystals.

In this work, we calculate the melting slope using a similar approach to that of Aguado and Madden.
We minimise finite-size effects and maximise the lengths of equilibration and production simulations by performing
molecular dynamics with a highly-accurate and sophisticated atomistic force-field\cite{long}.
We use perturbation theory to correct the small differences between our force-field's description of the potential
energy surface and that of DFT.
We also take precautions to ensure that the configurations generated by our force-field are very close to those that
would be generated directly from the DFT potential energy surface: we parametrize this model by fitting to DFT forces, stresses, and 
energies calculated on the hot solid {\em and the liquid}; we also perform a first principles molecular dynamics simulation of liquid MgO 
to verify that we get a liquid structure that is very similar to that produced by our force-field.

An important conclusion of the present work
is that, on the scale of the discrepancy between theory and experiment, 
 there is relative agreement between calculations of the DFT melting slope.
It is important to note that there have been DFT-based
calculations whose results differ strongly from those of Alf\'e, Aguado and Madden, and us.
However, we draw a distinction between DFT-based calculations that simply use models that 
rely to some extent on DFT, and calculations of the DFT melting slope. The latter are calculations which arrive at a
close approximation of the melting slope that would be obtained from large scale molecular dynamics simulations
on the DFT potential energy surface.
As an example of the former kind of calculation we consider the study by
Strachan {\em et al.} who calculated the melting curve using a model that 
was fit to the DFT equations of state (equilibrium volumes, bulk moduli, etc..) of the 
B1 and the high pressure B2 phases of MgO
as well as the shear stresses along the transformation path between them. Therefore, this fit was to a very small 
amount of {\em zero temperature} DFT data and the resulting force-field was used directly to compute the melting line. 

The fit of our models is to high temperature solid and liquid DFT data and we converge this fit with respect 
to the quantity of DFT data ($\sim 5,000$ numbers are typically required). The closeness of the fit\cite{long} and
the effectively-infinite amount of data used in the fit allows us to be confident that the force-field provides a very 
accurate description of the forces on the ions.
However, the crucial point, as explained below, is that the role of our model is only
to provide us with realistic statistically-independent hot solid and liquid configurations. 
The melting slope is computed by performing DFT calculations directly on these configurations so that, by first-order perturbation theory, 
we arrive at a close approximation to the DFT melting slope. 
Furthermore, we estimate the closeness of this approximation below.
So, despite the fact that we have a very accurate DFT-parameterised model, our goal is not to calculate 
this model's melting slope but to use it as a stepping stone to calculate
the DFT melting slope.

Semi-empirical DFT calculations were performed by Cohen and Gong\cite{cohen_and_gong}, however, their Potential-Induced-Breathing (PIB) 
model imposes unphysical constraints on the density. For example, it is known that oxygen ions are highly polarisable but, within their approach, 
they remain spherically-symmetric. This results, among other effects, in a vast over-estimation of 
longitudinal optical phonon frequencies
\cite{chizmeshya}. It is not known how oxygen polarisation affects the structure of liquid MgO, for example, but it is clear both from our classical
and from our {\em ab initio} molecular dynamics simulations that oxygen ions acquire large dipoles in the disordered solid and in the liquid.

\section{Calculations}
We determine the  melting slope $dT_m/dP$
of MgO at zero pressure by using the Clapeyron relation.
We calculate $\Delta V$ and $\Delta E$ with classical molecular dynamics and apply corrections to
them using DFT.

DFT calculations were carried out within the local 
density approximation (LDA) using norm-conserving pseudopotentials with and without 
core corrections~\cite{core_correction} for Mg and O, respectively, and a plane wave basis set 
with 120 Ry energy cut off. Simulation cells contained 64 atoms and the 
Brillouin zone was sampled with the $\Gamma$-point. Tests with 8 k-points 
yielded negligible ($<1$\%) differences on solid-liquid energy differences,
with respect to $\Gamma$-point sampling. 

In spite of their lower accuracy, model potentials can speed up 
considerably the task of calculating $\Delta V$ and $\Delta E$ from 
first principles if they are used as a ``reference'' model
for the first-principles potential~\cite{alfe}. The model potential is used to generate 
statistically significant atomic configurations at the P-T conditions of 
interest and the first-principles values of $\Delta V$ and $\Delta E$ are then
obtained by performing DFT calculations on those configurations only. 
We will show that, thanks
to the quality of the model potential used in this work, the errors introduced 
by this procedure are significantly smaller than those intrinsic to the standard 
approximations to DFT, which therefore remains the main source of uncertainty
in our calculations.
In order to achieve such a level of precision we use a model potential for MgO recently
developed by us, which accounts for arbitrary aspherical distortions of the oxygen 
valence shell~\cite{long}. Its parameters are obtained by best fit to DFT forces,
stresses and energy in atomic configurations which are representative of 
the physical conditions of interest~\cite{laio}. 
For this study we have used
one potential ($\Phi_l$) which was optimized in the liquid at 3000 K and $P=0$ GPa,
and another ($\Phi_s$) which was optimized at the same P-T conditions in the solid.
Average energies were set to be identical 
to the DFT values (this can trivially be imposed through an arbitrary additive
constant).  
For both potentials, phonons, thermal expansion,
and equations of state across a wide range of
temperatures and pressures, are in very good agreement with 
experiments and with independent DFT calculations\cite{long}.

Fig. 1 shows $\Delta V$ and $\Delta E$ as extracted from long ($\sim 100$ ps) molecular dynamics
simulations of the solid (with $\Phi_s$) and the liquid (with $\Phi_l$)
in a range of temperatures 
close to the experimental values for $T_m$ ($3040\pm 100$ K 
\cite{zerr_and_boehler} or $3250\pm20$ K \cite{ronchi}). Simulations
were performed in cells containing $512$ atoms under periodic boundary conditions. 
We verified that finite size effects on volume, compressibility,  and thermal expansivity 
were negligible with this cell size.

The first-principles values of $\Delta V$ and $\Delta E$ can be obtained by series 
expansion in the difference between the reference potential and the first-principles
potential {\bf \cite{alfe} }. The first-principles energy can be obtained as 
$E_{\mathrm{fp}} =  \langle \Phi_{\mathrm{fp}} \rangle_{\mathrm{fp}} = 
\langle \Phi_{\mathrm{mp}} \rangle_{\mathrm{mp}} +
\langle \Phi_{\mathrm{fp}} - \Phi_{\mathrm{mp}} \rangle_{\mathrm{mp}} + 
\beta \langle (\Phi_{\mathrm{fp}} - \langle \Phi_{\mathrm{fp}} \rangle_{\mathrm{mp}})
              (\Phi_{\mathrm{fp}} - \Phi_{\mathrm{mp}}) \rangle_{\mathrm{mp}}
+ O(\beta^2) $, 
where $\Phi_{\mathrm{mp}}$ is the model potential ($\Phi_s$ or $\Phi_l$), $\Phi_{\mathrm{fp}}$ the 
first-principles potential, $\beta = 1/k_BT$ ($k_B$ is the Boltzmann constant), 
and statistical averages on the model or first-principles potential are indicated
with $\langle ...\rangle_{\mathrm{mp}}$ and $\langle ...\rangle_{\mathrm{fp}}$, respectively.
Similarly, for the first-principles value of the molar volume we have,
to lowest order in $\beta$ and $\Delta V$:
$V_{fp} \simeq V_{\mathrm{mp}} + V_{\mathrm{mp}} K_T \langle P_{\mathrm{mp}}-P_{\mathrm{fp}} \rangle_{\mathrm{mp}}$, 
where $P_{\mathrm{fp},\mathrm{mp}}$ is the pressure calculated from first-principles or with
the model potential, and $K_T$ is the isothermal compressibility.
The first-principles value of $\Delta E$ ($\Delta V$) is then 
obtained as the difference between the values of $E_{\mathrm{fp}}$ ($V_{\mathrm{fp}}$) in
the liquid and in the solid.
 
We computed $E_{\mathrm{fp}}$ and $V_{\mathrm{fp}}$ at 3070 K and $P=0$ GPa
both in the liquid and in the solid
by using twenty statistically independent configurations extracted from a long molecular
dynamics run with the model potential. Each configuration was separated from the previous 
one by tens of picoseconds. Because the potentials have been arbitrarily 
given an energy offset so that $\langle \Phi_{\mathrm{fp}} - \Phi_{\mathrm{mp}} \rangle_{\mathrm{mp}} = 0$ at 3000 K,
the first significant term of the series expansion for the energy is the linear term
in $\beta$. We verified that this term is indeed very small (-4.3  K in the liquid and
-2.6 K in the solid), which implies that higher terms can be safely neglected. 

The same holds
true for the volume, where we find that $\langle P_{\mathrm{fp}}-P_{\mathrm{mp}} \rangle_{\mathrm{mp}} = 0.06$ GPa,
with a mean square deviation of 0.4 GPa, which means that uncertainties in 
the determination of the first-principles volumes are of the order of 1\% . 
 
The very good performance of our model potential on the thermal 
expansion for the solid phase \cite{long} suggests that the agreement 
on $\Delta V$ between DFT and the potentials in Fig. 1 can be extended to 
all temperatures in the vicinity of 3070 K. A similar conclusion can be reached
for the energy difference, based on the fact that energy fluctuations,
and therefore heat capacities \cite{textbook}, are correct to within 
$10\%$ \cite{long}. We can summarize the above considerations by saying that 
the data of Fig. 1 represent the first-principles values of $\Delta E$ within 10\%
and of  $\Delta V$ within 2 \%. 
\begin{figure}
\epsfig{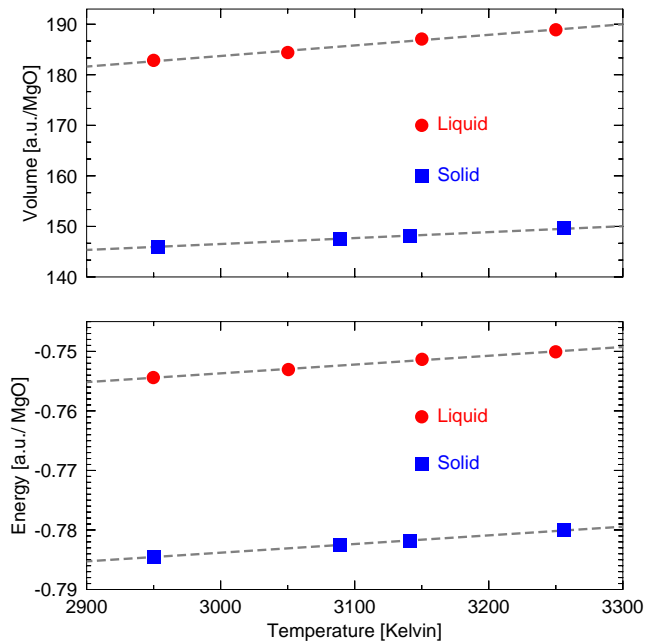}
\caption{Energy and volume of liquid and solid MgO as a function
of temperature from molecular dynamics simulations.}
\label{fig:fig1}
\end{figure}
\begin{figure}
\epsfig{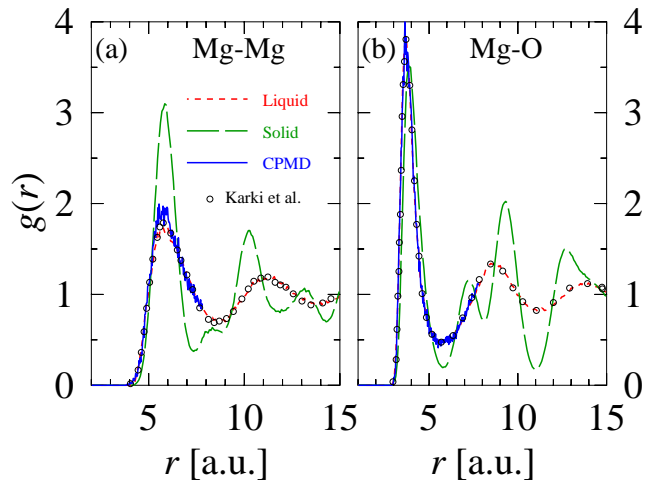}
\caption{Pair correlation functions of solid and liquid MgO at
$\sim 3100$ K from molecular dynamics using a model potential
compared to the results of first principles molecular dynamics of the liquid.
Both our Car-Parrinello molecular dynamics (CPMD) simulations 
and Born-Oppenheimer molecular dynamics simulations by 
 Karki {\em et al.}\cite{stixrude} are presented.}
\label{fig:fig2}
\end{figure}

What is clear from Fig. 1 is that neither  $\Delta V$ or $\Delta E$ is
strongly temperature dependent and that the melting slope, therefore, depends
approximately linearly on the melting temperature. The uncertainty in $T_{m}$ 
(and therefore $dT_m/dP$) is
of the order of $\sim 10\%$ which is much less than 
the discrepancy with experiment on $dT_m/dP$ ($> 300 \%$) which we want to address in this work.
Fits of straight lines to the data of Fig. 1 yield 
\begin{eqnarray}
\Delta E = 0.0295 + 1.97\times 10^{-7} T \nonumber \\
\Delta V =  9.71 +  9.16\times 10^{-3} T \label{onlyeqn}
\end{eqnarray}
From this we can calculate melting slopes
ranging from $dT_m/dP = 130$ K/GPa if  $T_m = 3050$K to $dT_m/dP = 145$ if $T_m = 3250$ K

For the sake of completeness, we have attempted a determination of
$T_m$. This was achieved by first calculating the melting 
temperature of $\Phi_l$ by means of the two-phase method\cite{belonoshko}, and then
correcting the results using perturbation theory \cite{alfe}.
A simulation cell containing $1024$ atoms was used for the two-phase
method. Previous 
investigations\cite{belonoshko,vocadlo_and_price}
have concluded that the finite size effects are negligible with this size.
We find that $T_m = 3010 \pm 50$ K for $\Phi_l$, the error reflecting only 
statistical uncertainties related to the two-phase method,
and not other systematic errors due to the optimization procedure or to
approximations to DFT. However, we caution that, in this simulation, potential $\Phi_l$
is used to describe both the liquid and the solid phases, at 
variance with the calculations of $\Delta V$  and $\Delta E$ where two
different potentials were employed to describe the liquid and the solid 
and which therefore resulted in a much more accurate calculation.
From \ref{onlyeqn}, therefore, we find that the value of the melting slope calculated with 
our model potential, before DFT corrections have been applied,  is $127$ K/GPa.
We can calculate DFT corrections to the Gibbs free energy difference
between solid and liquid using the method described in Ref.~\onlinecite{alfe}.
If we do so, we find that the DFT Gibbs free energy differences are $17\%$ larger than those
calculated with the model potential, indicating that
the model potential overstabilizes the liquid with respect to
the DFT potential. This brings about a similar correction 
for $T_m$ \cite{alfe}, so that the ``DFT'' value of $T_m$ is estimated 
to be about 3500 K, and the melting slope 156 K/GPa.

We have used DFT to correct the values of $\Delta V$ and $\Delta E$
obtained with our model potential and, as a result, can be confident
that we have calculated the free energy difference between the solid 
and the liquid generated from $\Phi_l$ with
very close to DFT accuracy. However, there remains the possibility that 
the structure of the liquid generated by $\Phi_l$ is not
realistic. Given the accuracy of this potential and that it has been 
parameterized from DFT calculations of the liquid, this possibility would
appear to require a liquid-liquid phase transformation 
that involves changes in the electronic structure of the ions
that cannot be captured by our model's phenomenological representation of the electrons.
It is important to consider this possibility, therefore, we have
performed Car-Parrinello molecular dynamics (CPMD) simulations\cite{cpmd} of $64$ atoms of
liquid MgO at $3050$ K. These simulations were performed at zero pressure using 
variable cell dynamics\cite{constantp}. A small fictitious mass of 
$\mu = 100$ a.u. was used and temperature and pressure were 
corrected as described in Ref.\onlinecite{us_cpmd}. The CPMD simulation was a continuation
of a very long simulation using potential $\Phi_l$. After $2$ ps of equilibration with 
CPMD, a $1.5$ ps production run was used to compute the radial distribution functions
of the liquid. The results are plotted in Fig.\ref{fig:fig2} and compared
with the results of Born-Oppenheimer molecular dynamics simulations carried out 
by Karki, Bhattarai, and Stixrude\cite{stixrude}. There is near-perfect agreement on the pair-distribution
functions between the two independent first-principles simulations and our simulations of
a $512$-atom supercell using our model potential, $\Phi_l$. The agreement between the first principles
simulations seems to rule out major structural artefacts of the starting configuration in both
simulations.  The agreement with the structure of the liquid generated with $\Phi_l$ confirms
the reliability of this potential and also appears to rule out large finite-size effects on
the pair-distribution functions. The results presented in Fig.\ref{fig:fig2} are strong evidence that
realistic liquid structures have been used to calculate DFT corrections to $\Delta V$ and $\Delta E$
and, therefore, that we have calculated these quantities with very close to DFT accuracy.

It is important to note at this point that our reported DFT results
have been obtained 
within the LDA, as this approximation has proven to be very accurate
in describing low temperature properties. 
However, non local corrections to the LDA, such as those contained in
generalized gradient approximation theories (GGA), are known to have
a significant effect on melting temperatures\cite{alfe_mgo,alfe2} 
Therefore, as a test of the importance of exchange and correlation effects we 
repeated the analysis of energy differences with a GGA functional \cite{pbe}.
We find that average GGA energy fluctuations at $3000$ K 
are within 12\% of those calculated with the model potential.
Moreover, we find that the correction 
to the Gibbs liquid-solid free energy difference is only 2.7\%, which implies
a value for the GGA $T_m$ of 3090 K. This improves dramatically 
the agreement of $T_m$ with experiment, with respect to the LDA, 
and confirms that exchange and correlation effects are indeed important
in the determination of $T_m$.  
We caution, however, that 
the potential was constructed by fitting LDA quantities, so it is possible 
that the atomic configurations chosen for the comparison were not fully
representative of the GGA potential. 
We did not attempt a determination of $\Delta V$ with GGA as this would require a very expensive 
equilibration with the GGA functional. 
$\Delta V$ would have to be an order of magnitude 
larger than the LDA value to resolve the discrepancy with experiment on
the melting slope, which is highly unlikely.

Alf\'{e} has pointed out\cite{alfe_mgo,alfe_personal} that, because the 
energy gap between occupied and unoccupied electronic states is significantly
smaller in the liquid than in the solid, it is more appropriate 
to perform DFT calculations with a finite electronic temperature.
This results in a lowering of $T_m$ by approximately $500$K. 
Consideration of this correction brings our LDA and GGA values of $T_m$ into
very good agreement with those of Alf\'{e}.
Alf\'{e} reports that the correction to $\Delta E$ from this effect is almost $0.0036$ a.u.
per molecular unit. Because energy gaps calculated within LDA are generally too small (by as much as $\sim 50\%$), 
the error incurred in $\Delta E$ is likely to be significantly smaller than
this. In any case, for a fixed value of $T_m$, the reduction of $\Delta E$ by inclusion
of the electronic entropy contribution to the free energy should {\em increase} the calculated
melting slope thereby bringing the calculated value even further from the experimental 
value of Zerr and Boehler.

\section{Conclusions}

To summarize, we find that that the DFT/LDA value for the melting slope of
MgO ranges from $\sim 130$ K/GPa to $\sim 150$ K/GPa, depending primarily on the value chosen
for $T_m$ and with an overall uncertainty of about 10-15\% due to the model 
potential and to statistical sampling.  
We can safely conclude that the DFT/LDA result
is a factor of 3 to 4 larger than the value of $36$ K/GPa found in Zerr and Boehler's
experiment. 

There remains a small difference in the melting slopes calculated here and by Alf\'{e}.
The source of this difference is unknown, but may be due to differences in the details of the DFT calculations
such as our different pseudopotential representations of the inert core electrons.
The important point is that this difference is substantially smaller than the discrepancy with the experimental
results of Zerr and Boehler. The present work, when taken together with the results of Alf\'e and
Aguado and Madden, shows that on the scale of the discrepancy between calculations
and experiment, there appears now to be a convergence of the results of calculations of the melting slope.
The only major source of error that would effect all of these calculations is the approximation to the
exchange-correlation energy.
However, tests with a GGA exchange-correlation functional show that, although the choice of
functional changes the values of $T_m$ and $dT_m/dP$, these changes 
are relatively small. Therefore it
seems highly unlikely that inadequacies of the DFT approximations used can fully
explain the historical discrepancy between theory and experiment.

The possibility of problems with the experiment
of Zerr and Boehler\cite{zerr_and_boehler}
have previously been suggested\cite{cohen_and_weitz,belonoshko,belonoshko_lif} and 
the fact that there is a large disagreement between experimental measurements of both
the melting temperature\cite{zerr_and_boehler,ronchi} and the melting slope\cite{zerr_and_boehler,zhang_and_fei}
suggests that more experimental work is necessary.
A determination
of the density change and latent heat at zero pressure are
crucial to resolve the issue. 
The disagreement between {\em our} DFT results and experiment could also be explained 
by assuming that the slope of the melting curve is initially 
very steep, but that it flattens out very
quickly, perhaps due to a liquid structure which changes rapidly under 
pressure to being much more similar to the solid. However, this explanation would 
not be compatible with the findings of Alf\'{e}, or Aguado and Madden, who explicitly
computed the melting temperature at high pressures. Therefore, we must discount this
possibility.

We note that the suggestion by Aguado and Madden that the discrepancy between theory
and experiment can be explained by the existence of a solid phase with a lower
free energy than the rocksalt structure cannot be correct. This is because, among possible 
solid phases, the one with the lowest Gibbs free energy has the highest melting temperature.
Therefore, the melting curve for a more stable solid phase should lie above the rocksalt
melting curve, i.e. it should have a higher $T_m$ at every pressure $P$.

The geophysical implications of a steep melting slope for MgO are
manyfold  \cite{alfe_mgo,zhang_and_fei}.
A steep melting slope implies that the melting temperature of (Mg,Fe)O,
of which MgO is an
end-member, is likely to be substantially higher than the geotherm and
comparable to the
melting temperature of (Mg,Fe)SiO$_3$ at lower mantle conditions.
This suggests, among other consequences, that large scale melting may
never have occurred in
the mantle~\cite{boehler_review}.\\

\section{Acknowledgments}
P.T. acknowledges support from the European Commission within the Marie Curie Support for Training and Development
of Researchers Programme under Contract No. MIRG-CT-2007-208858.

\end{document}